\documentclass[12pt]{article}
\begin{document}
\date{\mbox{ }}
\title{
\textbf{Anomaly for Model Building}\\
[5mm]}
\author{Utpal Sarkar\\
\\
{\normalsize \it Physical Research Laboratory, Ahmedabad 380 009,
India} } \maketitle

\thispagestyle{empty}

\begin{abstract}
\noindent A simple algorithm to calculate the group theory
factor entering in anomalies at four and six dimensions for
SU(N) and SO(N) groups in terms of the Casimir invariants of their
subgroups is presented. Explicit examples of some of the lower dimensional
representations of $SU(n), n \leq 5$ and SO(10) groups are
presented, which could be used for model building in four and six
dimensions.

\end{abstract}
\newpage
\baselineskip 18pt

The consistency of any gauge theory requires that the sum of
anomalies \cite{abj} due to all the fermions present in the theory
should cancel. The anomaly cancellation is necessary because any
classical symmetry is broken by quantum effects in the presence of
anomaly. In other words, any gauge theory with non-vanishing
anomaly is non-renormalizable \cite{anom2}.

In the standard model of electroweak interactions, the fermion
content is just right to make the theory anomaly free. In any
extensions of the standard model, particularly the ones which
introduces additional gauge symmetry, the most severe constraints
come from the anomaly cancellation conditions. Anomaly plays
crucial roles in theories of dimension higher than four. Since all
representations in odd dimensions contains both left and right
chiral fields, there is no anomaly. But in dimensions six, eight
or ten, any theory has to cancel higher dimensional anomalies. In
higher dimensional theories if the space is compactified in an
orbifold, then the orbifold compactification also lead to anomaly
at the fixed points. Consistency of such theories then require
that the brane anomalies at the fixed points should vanish.

Fermions in the loop contributes to anomaly. So, in
non-supersymmetric theories the fermion representations are
constrained by anomaly. But in supersymmetric theories, any chiral
superfield would contribute to anomalies. So, the superfields
containing both the scalar and fermion representations are
constrained by the anomaly cancellation requirement.

The cancellation of anomaly is thus an integral part of
constructing any consistent model in four or higher dimensions. It
is thus important to know the group theory factor of any
representation contributing to anomaly in four or higher
dimensions. In four dimensions one needs to calculate the triangle
anomaly, while at six dimensions it is a box anomaly and at eight
dimensions it is pentagon anomaly. These group theory factors for
higher groups become difficult to calculate. There are 
rigorous methods for searching anomanly free theories
that are used usually \cite{oldan}. In this article we
present a simple algorithm to calculate the group theory factor
appearing in the expression for anomaly in four and six dimensions
for Lie groups, which could be useful for model building. 

In four dimensions, if the fermions belong to a representation ${\cal R}$ of
${\cal G}$, then the group factor in the expression for anomaly
can be written in terms of the generators $T^a({\cal R}_1)$ for
this representation ${\cal R}_1$ of ${\cal G}$. Then the
contributions of fermions in a representation ${\cal R}_r$ to
anomaly will be proportional to
\begin{equation}
{}_3{\cal A} = {\rm tr}~ [ T^a ({\cal R}_r) T^b ({\cal R}_r) T^c
({\cal R}_r)] .
\end{equation}
We shall use the notation ${}_n{\cal A}$ to represent anomalies, so that
triangle anomaly in four dimensional theories is represented
by $n=3$; box anomaly in six
dimensions with $n=4$ and pentagon anomaly in eight dimensions by
$n=5$.

In general, $n = d/2 +1 $ for anomaly in a d-dimensional theory
and the group factor for fermions or chiral superfields in a
representation ${\cal R}_r$ entering in the expression for anomaly
is given by
\begin{equation}
{}_n{\cal A} = {\rm tr}_{{\cal R}_r} ~T^n = {\rm tr}~ [T^{a1}
({\cal R}_r) T^{a2} ({\cal R}_r) ~ \cdot \cdot \cdot ~T^{an}
({\cal R}_r)] .
\end{equation}
In four dimensions this anomaly factor for all the fermions or all
the superfields should cancel for consistency. In higher
dimensional theories one should apply this factor with caution.

Consider a six dimensional orbifold model
\cite{buch,hebe}, compactified on ${\cal
R}^4 \times T^2/Z_2$. There will be four dimensional anomaly at
the fixed points and also the six dimensional anomaly at the bulk.
If the gauge group in the bulk is $G$ and at the fixed points only
the group $H$ acts, then the only non-vanishing anomaly at the
fixed points will be restricted to the subgroup $H$ of $G$. The
brane anomalies are also associated with the parities that acts on
the fields due to the action of the discrete $Z_2$ symmetry. So,
the choice of parity to break $N=2$ supersymmetry to $N=1$
supersymmetry implies that the spinor in a vector superfield and
spinor in a scalar superfield contribute to anomalies with
opposite sign. There is another difference in anomalies at six
dimensions. In some cases there are contributions of the form
$$ A_{red} = ({\rm tr}_{R_r} ~T^2)^2 $$
in addition to the usual ${\rm tr}_{R)r} ~T^4$ terms. This
factorized contributions, also known as reducible anomalies, does
not have any analogy in four dimensions and could be cancelled by
introducing antisymmetric tensor fields and utilizing the
Green-Schwarz mechanism. We shall thus calculate the group factor
for the irreducible anomaly contributions, given by ${}_4{\cal
A}$. Since there are no independent 4th order invariants for the
groups SU(2) and SU(3) and ${\rm tr}_{R_r} T^4 = ({\rm tr}_{R_r}
T^2)^2$, we shall not present ${}_4{\cal A}$ for these two groups.

In the present approach all invariants are calculated in terms of
invariants of the subgroups. Consider the subgroup
\begin{equation}
{\cal G}_1 \times {\cal G}_2 \subset {\cal G},
\end{equation}
where ${\cal G}_2 = U(1)$ is an abelian subgroup of ${\cal G}$. We
can then decompose any representations of ${\cal G}$ under ${\cal
G}_1 \times {\cal G}_2$ subgroup as
\begin{equation}
{\cal R} = \sum_i (r_i, f_i)
\end{equation}
where $f_i$ are the $U(1)$ quantum numbers of ${\cal G}_2$. For
any group $SU(m)$ we can write down the decomposition of the
fundamental $m$ dimensional representation under the subgroup
$SU(m-1) \times U(1)$ as
\begin{equation}
m = (m-1,1) + ( 1, -m+1).
\end{equation}
The second numbers $1$ and $(-m+1)$ are the $U(1)$ quantum numbers
in this decomposition. Using the product decomposition formulas we
can find out the decomposition of all other representations of
$SU(m)$ under the subgroup $SU(m-1) \times U(1)$.

We can then use the formulas
\begin{eqnarray}
  {}_3 {\cal A} ({\cal R}) &=& \sum_i T^2(r_i) \cdot f_i \nonumber \\
  {}_4{\cal A} ({\cal R}) &=& \sum_i {}_3{\cal A} (r_i) \cdot f_i \nonumber \\
  {}_n{\cal A} ({\cal R}) &=& \sum_i {}_{(n-1)}{\cal A} (r_i) \cdot f_i
\end{eqnarray}
to calculate the anomalies for the representation ${\cal R}$ of
the group ${\cal G}$ in terms of the invariants of the subgroup
${\cal G}_1 \times {\cal G}_2$. For verification of the results we
also use the formulas
\begin{equation}
{}_n{\cal A} ({\cal R}) = \sum_i {}_n{\cal A} (r_i) .
\end{equation}
For completeness we also present a couple of useful relations
\begin{eqnarray}
{}_n {\cal A} ( {\cal R}_1 + {\cal R}_2 ) &=& {}_n {\cal A} (
{\cal R}_1 ) + {}_n {\cal A} ( {\cal R}_2 ) \nonumber \\
{}_n {\cal A} ( {\cal R}_1 \times {\cal R}_2 ) &=& {}_n {\cal A} (
{\cal R}_1 ) D({\cal R}_2) + {}_n {\cal A} ( {\cal R}_2 ) D({\cal
R}_1) .
\end{eqnarray}
Thus by writing down the decomposition of any representation under
its subgroup containing a $U(1)$ factor, it will be possible to
calculate the irreducible group factor entering in the expression
for anomalies. For the cancellation of anomalies in any theory, we
require only this factor and hence these results will be extremely
useful while building models in many extensions of the standard
model.

\begin{table}[h!]
\caption{Quadratic Casimir invariant and triangle anomalies of
SU(3). There are also no 4th order invariants.}
\begin{center}
\begin{tabular}{ccc}
\hline \hline
${\cal R}_r$ & Index (l) & ${}_3 {\cal A}$ \\
\hline &&\\
3 & 1 & 1 \\
$\bar{3}$ & 1 & $-1$  \\
6 & 5 & 7 \\
$\bar{6}$ & 5 & $-7$ \\
8 &6 &  0 \\
10 & 15 & 27 \\
$\bar{10}$ & 15 & $-27$ \\
15 & 20 & 14 \\
$15^\prime$ & 35 & 77 \\
21 & 70 & $-182$ \\
24 & 50 & $-64$ \\
27 & 54 & 0 \\
&&\\
\hline \hline
\end{tabular}
\end{center}
\end{table}

Let us first consider the group SU(2). Since all representations
are pseudo-real, all four dimensions triangle anomalies vanishes.
There are also no fourth order invariants and hence we have to
worry only about the quadratic Casimir invariants. For the group
SU(2) the quadratic Casimir invariants are given by
\begin{equation}
T^2 (N) = \sum_{i=-(N-1)/2}^{(N-1)/2} |i|^2
\end{equation}
for an $N$ dimensional representation.

\begin{table}[t!]
\caption{Anomalies for the group SU(4). }
\begin{center}
\begin{tabular}{cccc}
\hline \hline
${\cal R}_r$ & Index (l) & ${}_3 {\cal A}$  & ${}_4 {\cal A}$ \\
\hline &&&\\
4 & 1 & 1 & 1 \\
$\bar{4}$ & 1 & $-1$ & 1 \\
6 & 2 & 0 & $-4$ \\
10 & 6 & $8$ & 12 \\
$\bar{10}$ & 6 & $-8$& 12 \\
15 & 8 & 0 & 8 \\
20 & 13 & $-7$ & $-11$ \\
$\bar{20}$ & 13 & 7 & $-11$ \\
$20^\prime$ & 16 & 0 & $-56$ \\
$20"$ & 21 & $-35$ & 69 \\
35 & 56 & 112 & 272 \\
36 & 33 & 21 & 57 \\
45 & 48 & 48 & 24 \\
50 & 70 & 0 & $-380$ \\
&&&\\
\hline \hline
\end{tabular}
\end{center}
\end{table}

We shall next consider the group SU(3). The fourth order
invariants are again absent and hence we have to compute the
quadratic Casimir invariants and the triangle anomalies for the
different representations. Using the decompositions of the
representations of SU(3) under $SU(2) \times U(1)$ as
\vbox{
\begin{eqnarray}
3 &=& (2,1) + (1,-2) \nonumber \\
\bar{3} &=& (2,-1) + (1,2) \nonumber \\
6 &=& (3,2) + (1,-4) + (2,-1) \nonumber \\
&& \cdot \cdot \cdot ~~~ \cdot \cdot \cdot \nonumber
\end{eqnarray}}
we can use equations 6 and 7 to calculate the quadratic Casimir
invariants and the anomalies, which is presented in table 1.
The triangle anomalies for SU(3) can be computed using the
formula presented in ref. \cite{su3}. 

Proceeding in the similar way, we can calculate the triangle and
box anomalies for the groups SU(4) and SU(5), which are presented
in tables 2 and 3. The triangle anomalies can again be computed
and compared following ref. \cite{su3}. Calculating the box
anomalies are more involved. In general, the relations involve
both reducible as well as irreducible anomalies. For the group 
$SU(n) (n>3)$ the anomaly for the adjoint representation can be 
written as
\begin{equation}
{}_4{\cal A} (adj) = 2 ~ n ~ {}_4{\cal A} (fund) + 6 ~ 
({\rm tr}_{(fund)} T^2)^2
\end{equation}
in terms of the invariants of the fundamental representations. 
However for purpose of anomaly cancellation in six dimensional
theories we are interested in only the irreducible anomalies
and hence the present method will serve the purpose. 
This procedure can be extended to any higher
groups in the same way.

\begin{table}[t!]
\caption{Anomalies for the group SU(5).}
\begin{center}
\begin{tabular}{cccc}
\hline \hline
${\cal R}_r$ & Index (l) & ${}_3 {\cal A}$ & ${}_4 {\cal A}$ \\
\hline &&&\\
5 & 1 & 1 & 1 \\
$\bar{5}$& 1&$-1$&1 \\
10 & 3 & 1 & $-3$ \\
$\bar{10}$ &3 &$-1$& $-3$ \\
15 & 7 & 9 & 13 \\
$\bar{15}$ & 7 & $-9$& 13\\
24 & 10&0&10 \\
35 & 28 & $-44$& 82\\
$\bar{35}$&28&44&82\\
40&22&$-16$&$-2$\\
45&24&$-6$&$-6$\\
50&35&$-15$&$-55$\\
70&49&29&79\\
$70^\prime$&84&$-156$&354\\
75&50&0&$-70$\\
&&&\\
\hline \hline
\end{tabular}
\end{center}
\end{table}

We shall now consider a slightly non-trivial case of the group
SO(10). We consider the decomposition of SO(10) under the subgroup
$SU(5) \times U(1)$. The vector and the spinor representations of
SO(10) decompose under $SU(5) \times U(1)$ as
\begin{eqnarray}
10 & = & (5,2) + (\bar{5},-2) \nonumber \\
16 &=& (1,-5) + (\bar{5},3) + (10,-1). \nonumber
\end{eqnarray}
If we now calculate the triangle anomalies in terms of the
triangle anomalies of SU(3) representations, then it is obvious
that both the representations 10 and 16 have vanishing anomalies,
since the SU(5) anomalies ${}_3{\cal A} (5) = {}_3{\cal A} (10) =
- {}_3{\cal A} (\bar{5}) = 1$. It is also well-known that all
representations of $SO(10)$ group are anomaly-free. Using equation
6 we can calculate the box anomalies for the representations of
SO(10), which are given by
\[
\begin{array}{ccccccccc}
{\cal R} & \rightarrow & 10&16&45&54&120&126&210\\
{}_4{\cal A} &\rightarrow&4&-4&8&72&-8&-104&120\\
\end{array}
\]
modulo a normalization factor, which will not change the condition
for anomaly cancellation. From these the two relations follow:
\[
{}_4{\cal A} (45) = 2~ {}_4{\cal A} (10) ~~~ {\rm and} ~~~
{}_4{\cal A} (16) = {}_4{\cal A} (\bar{16}) = - {}_4{\cal A} (10).
\]
These relations are true only for the irreducible anomalies, as we
discussed earlier. This method can be extended to higher dimensional
theories and to all the Lie groups. 

In summary, we presented a simple algorithm of calculating anomalies at
four and six dimensions for all the Lie groups. We gave explicit
example for the groups SU(3), SU(4), SU(5) and SO(10), which are
extensively used in building orbifold grand unified theories and 
Higgsless models in six dimensions and extensions
of the standard models.


\end{document}